\documentclass[twocolumn,showpacs,preprintnumbers,amsmath,amssymb]{revtex4}

\usepackage{graphicx}
\usepackage{dcolumn}
\usepackage{bm}
\def\eion{{$(e~+~ion)$}\ }
\def\etal{{\it et\thinspace al.}\ }
\def\fexvii{{\rm Fe {\sc xvii}}\ }
\def\fexviii{{\rm Fe {\sc xviii}}\ }
\def\en{{$n$\ }}

\begin{document}


\title{Highly Excited Core Resonances in Photoionization of \fexvii :
Implications for Plasma Opacities}
%

\author{Sultana N. Nahar$^1$, Anil K. Pradhan$^1$,
Guo-Xin Chen$^{2}$, Werner Eissner$^3$}
\email{nahar@astronomy.ohio-state.edu}
\affiliation{$^1$Department of Astronomy, The Ohio State
University,Columbus, Ohio
43210,$^2$ ITAMP, Harvard-Smithsonian Center for Astrophysics, 60
Garden Street, Cambridge, MA 02138, 
$^3$Institut f\"ur Theoretische Physik, Teilinstitut 1, 70550
Stuttgart, Germany}

\date{\today}

\begin{abstract}
A comprehensive study of high-accuracy photoionization cross sections is
carried out using the relativistic Breit-Pauli R-matrix (BPRM) method
for (h$\nu$~+~\fexvii $\rightarrow$ \fexviii + e). Owing to its importance
in high-temperature plasmas the calculations cover a large energy range,
particularly the myriad photoexciation-of-core (PEC) resonances including 
the $n$ = 3 levels not heretofore considered. The calculations employ a 
close coupling wave function expansion of 60 levels of the core
ion \fexviii ranging over a wide energy range of nearly 900 eV
between the \en = 2 and \en = 3 levels. Strong coupling effects
due to dipole transition arrays $2p^5 \rightarrow 2p^4 \ (3s,3d)$  
manifest themselves as large PEC resonances throughout this range, and 
enhance the effective photoionization cross sections orders of magnitude 
above the background. Comparisons with the erstwhile Opacity Project (OP) 
and other previous calculations shows that the currently available cross
sections considerably underestimate the bound-free cross sections. A
level-identification scheme is used for spectroscopic designation of the
454 bound fine structure levels of \fexvii, with $n \leq 10$, $l \leq$ 9, 
and 0 $\leq J \leq$ 8 of even and odd parities, obtained using the ab 
initio BPRM method (compared to 181 LS bound states in the OP work). The 
calculated energies are compared with those available from the National 
Institute for Standards and Technology database, which lists 63 levels
with very good agreement. Level-specific photoionization cross sections 
are computed for all levels. In addition, partial cross sections for 
leaving the core ion \fexvii in the ground state are also obtained. These 
results should be relevant to modeling of astrophysical and laboratory 
plasma sources requiring (i) photoionization rates, (ii) extensive 
non-local-thermodynamic-equilibrium models, (iii) total unified 
electron-ion recombination rates including radiative and dielectronic 
recombination, and (iv) plasma opacities. We particularly examine PEC 
and non-PEC resonance strengths and emphasize their expanded role to
incorporate inner-shell excitations for improved opacities, as shown by 
the computed monochromatic opacity of \fexvii. 
\end{abstract}

\maketitle

\section{\label{sec:level1}{Introduction}}
High precision studies of photoionization of atoms in a large number of
excited levels are of interest in several areas. Derivative quantities 
such as laboratory and astrophysical plasma opacities 
\cite{bailey,symp,op,opal}, spectral models of photoionization 
dominated soures such as active galactic nuclei \cite{grupe},
non-local-thermodynamic-equilibrium models of stellar atmospheres
\cite{mihalas}, and total electron-ion recombination rates \cite{np04}, 
depend on the accuracy of underlying treatment of photoionization for 
the {\it entire} atom(ion) at all energies where it is abundant under 
speciffic plasma conditions.

There have been a number of theoretical calculations to produce large
amounts of data for practical applications, in particular under the
Opacity Project (OP) \cite{op} and the Iron Project (IP) \cite{ip}.
But they are usually limited in terms of accuracy owing to computational
constraints, and in detailed
examination of the photoabsorption process for excited levels.
Furthermore, except for light elements or simple few-electron systems,
large-scale calculations do not adequately treat
the two most important atomic effects,
relativistic fine structure and resonances.
Many previous calculations are
basically in LS coupling \cite{op}, or which take account of fine structure
through an algebraic transformation rather than a relativistic
calculation per se. The primary reason of course  is the difficulty of
doing so, since the calculations for photoionization of a large number
of fine structure levels remains an enormous task. Close coupling
calculations that account for the ubiquitous resonances in
photoionization
cross sections are mostly done using the R-matrix method
\cite{br75,burke,bb93}.  The
extension to include relativistic effects led to the development of the
Breit-Pauli R-matrix (BPRM) codes \cite{82s1,ben95},
and to aforementioned work under the Iron
Project for relatively simple atomic systems such as the He-like and
Li-like ions (e.g. \cite{np06}).

 Such large-scale calculations with high precision may be of crucial
importance in resolving a major astrophysical problem \cite{asp09} 
that is related to
plasma opacities \cite{pn09} and recent studies of solar
abundances of common elements such as C, N, O and Ne. The new abundances
are widely discordant with the {\em standard} solar abundances by up to
50\% \cite{asp09}. The new abundances are derived from high-resolution
spectroscopy and the most advanced
3D hydrodynamic non-local-thermodynamic-equilibrium (NLTE) models
However, such a huge decrease in solar
elemental composition is not supported by heliosmological
observations of solar oscillations measured with great accuracy
\cite{ba08}. In addition, predictions of solar interior models are also
in conflict with the new abundances \cite{fd06}. Given the inverse
relationship between opacity and abundances, tt has been suggested
that a marginal, but highly significant, increase in mean opacities of
about 10-20\% could possible reconcile the new and the standard solar
abundances \cite{bahcall05}.

 Another need for higher precision opacities arises due to the recent
capability of creating stellar interior conditions in the laboratory
\cite{bailey}. Inertial confinement fusion (ICF) devices can now produce
plasmas in LTE at temperatures and densities approaching those at the
boundary of the radiative zone and the convection zone (CZ) in the Sun.
Measurements have been carried out at the 
Z-pinch facility at the Sandia National Laboratory 
of monochromatic opacity spectra of iron at T$_e \sim$ 150-200 eV 
and N$_e \sim 10^{22-23}$/cc. \fexvii is the largest
contributor to opacity amongst the iron ions prevalent
at that temperature and density, from {\rm Fe~{\sc xvi}} - {\rm Fe~{\sc
xix}} with ionization fractions 0.0.0757, 0.302, 0.366 and 0.18
respectively. However, there are marked differences
between the Z-pinch observations and theoretically
computed opacities \cite{bailey}. Thus, 
both astrophysical and
laboratory observations require high-precision studies
to understand or resolve the discrepancies.

We now note one of the basic atomic physics assumptions made in earlier
theoretical calculations of plasma opacities: {\it resonances are treated 
as lines}, and perturbative treatments are then employed to include 
continuum contribuions. Depending of the temperature and density, most 
of the opacity could stem from inner-shell excitations in complex atomic 
species such as iron ions with closed L and M subshells \cite{symp,sb04}. 
Such {\em photoexcitation-of-core} (PEC) process occurs at high
energies above the first ionization threshold. Therefore, it manfiests
itself as resonances in the bound-free. The corresponding PEC resonances 
are large features present at incident photon frequencies corresponding to
dipole transitions in the core ion \cite{ys87,pn11}. It follows that
the major source of opacity in {\em any} plasma source, astrophysical or in the
laboratory, lies in bound-free resonances in general, 
heretofore treated mostly as
bound-bound transitions or lines, and in PEC resonances in particular. 
But the approximation of resonant photoionization by lines was made in the
OP \cite{symp,op} and OPAL\cite{opal} projects because opacities calculations
are extremely intensive, and require huge amounts of atomic
data for all consituent elements in many excitation/ionization states.  
Nevertheless, good agreement is found between the two projects for
{\em mean} opacities over a wide range of temperatures and densities
(\cite{sb04}, discussed later).
But now, owing to the major problems outlined above, it appears necessary to
compute atomic parameters to higher precision 
to address this issue, albeit on a
relatively small scale for individiual ions to begin with. 

This report presents pilot and prototypical
calculations required for revised opacities work. Ne-like \fexvii is 
a prime contributor to solar
opacity near the convection zone boundary \cite{bailey}, with many
PEC features in photoionization cross sections. By carrying out
one of the most comprehensive BPRM calculations for
photoionization of an atomic system, we have computed and analyzed
level-specific fine structure cross sections, and the {\it distribution}
of resonances over a wide energy range encompassing 
complex atomic structures. 
  In previous studies of photoionization including resonances, 
\fexvii was studied mainly in the
low energy region using a small wavefunction representation. An R-matrix
calculation in LS coupling using a 2-state wave function expansion for
the core ion \fexviii was carried out under the Opacity Project
(M. P. Scott, see \cite{top}).
A later coupled channel calculation, aimed at unfied electron-ion
recombination calculations for (e~+~\fexviii) $\rightarrow$ \fexvii, 
employed the (BPRM) method using a 3-level (or 3CC) wave function expanion 
by \cite{pnz}. The recombination
spectrum obtained using 3CC photoionization compared very well with
the experimentally observed spectrum in the very low energy region. But
due to the ion's abundance in high temperature plasmas, a preliminary
calculation using a 60-level expansion for \fexviii was initiated by
\cite{znp}. This initial study on photoionization cross
sections focused only on one symmetry: levels with total $J$=0. 
But to examine its opacity {\it in toto} a much more extensive
calculation is necessary. This study is
a full-scale photoionization calculation using the large
60-level wave function expansion for \fexviii, with energy
levels up to approximately
1 keV. That energy range should account for nearly all of the resonance
structures due to L-shell excitation of \fexvii at temperatures
$\sim$200 eV where it is abundant near the solar CZ; the Planckian
blackbody distribution decreases to render \fexvii ionization fraction
and opacity to be insignifiant beyond E $\sim$ 1.5 keV. We
especially aim to elucidate photoexcitation-of-core
(PEC) resonance features that greatly affect photoionization in the high
energy region.

In this report we point out other potential applications as well.
 Since the cross sections 
are computed up to levels with $n(SLJ) \leq 10$, they should be 
useful for non-LTE spectral 
modeling in the X-ray region. The present work yields accurate
energies and spectroscopic identification of the 454 bound levels of
\fexvii computed through ab initio calculations. 
In addition, level-specific {\it partial photoionization cross sections}
of \fexvii into the ground state of \fexviii are computed, as 
required for future studies of unified recombination cross sections 
including both radiative and dielectronic recombination processes.

\section{\label{sec:level1}Theory\protect\\ 
}

Photoionization calculations have been carried out in the close 
coupling (CC) approximation using the R-matrix method as developed 
under the Opacity Project \cite{symp,op} and the Iron Project
\cite{ip,ben95}.
In the CC approximation 
the atomic system is represented as the 'target' or the 'core' ion of 
N-electrons interacting with the (N+1)$^{th}$ electron. The (N+1)$^{th}$ 
electron may be bound in the electron-ion system, or in the electron-ion
continuum depending on its energy to be negative or positive. The total 
wavefunction, $\Psi_E$, of the (N+1)-electron system in a symmetry 
$J\pi$ is an expansion over the eigenfunctions of the target 
ion, $\chi_{i}$ in specific state $S_iL_i(J_i)\pi_i$, coupled with the 
(N+1)$^{th}$ electron function, $\theta_{i}$:
\begin{equation}
\Psi_E(e+ion) = A \sum_{i} \chi_{i}(ion)\theta_{i} + \sum_{j} c_{j} \Phi_{j},
\end{equation}
where the sum is over the ground and excited states of the target or the
core ion. The (N+1)$^{th}$ electron with kinetic energy $k_{i}^{2}$ 
corresponds to a channel labelled $S_iL_i(J_i)\pi_ik_{i}^{2}\ell_i(SL(J)\pi)$. 
The $\Phi_j$s are bound channel functions of the (N+1)-electron system 
that account for short range correlation not considered in the first
term and the orthogonality between the continuum and the bound electron 
orbitals of the target.

The relativistic Hamiltonian in the Breit-Pauli R-matrix (BPRM)
approximation is given by
\begin{equation}
\begin{array}{l}
H_{N+1}^{\rm BP} = \\ \sum_{i=1}\sp{N+1}\left\{-\nabla_i\sp 2 - \frac{2Z}{r_i}
+ \sum_{j>i}\sp{N+1} \frac{2}{r_{ij}}\right\}+H_{N+1}^{\rm mass} + 
H_{N+1}^{\rm Dar} + H_{N+1}^{\rm so}.
\end{array}
\end{equation}
where the last three terms are relativistic corrections, respectively:
\begin{equation} 
\begin{array}{l}
{\rm the~mass~correction~term},~H^{\rm mass} = 
-{\alpha^2\over 4}\sum_i{p_i^4},\\
{\rm the~Darwin~term},~H^{\rm Dar} = {Z\alpha^2 \over 4}\sum_i{\nabla^2({1
\over r_i})}, \\
{\rm the~spin-orbit~interaction~term},~H^{\rm so}= Z\alpha^2 
\sum_i{1\over r_i^3} {\bf l_i.s_i}.
\end{array} 
\end{equation}

 Eq.(3) representes the one-body terms of the Breit interaction. Another
version of "full" BPRM codes has been developed including the two-body
terms, but those are only of importance in forbidden (E2, M1, M2)
transitions where two-electrons correlations may play an important role 
\cite{n10}. In contrast,
plasma opacities {\it in toto} are determined by strong E1 (dipole
allowed and intercombination) considered in this work.

Substitution of $\Psi_E(e+ion)$ in the Schrodinger equation
\begin{equation}
H_{N+1}\mit\Psi_E = E\mit\Psi_E
\end{equation}
introduces a set of coupled equations that are solved using the R-matrix 
method. The solution is a continuun wavefunction $\Psi_F$ for an electron 
with positive energies (E $>$ 0), or a bound state $\Psi_B$ at a 
{\it negative} total energy (E $\leq$ 0). The complex resonance structures 
in photoionization cross sections result from channel couplings between 
the continuum channels that are open ($k_i^2~>$ 0), and ones that are 
closed ($k_i^2~<$ 0). Resonances occur at electron energies $k_i^2$ 
corresponding to autoionizing states belonging to Rydberg series, 
$S_iL_i\pi_i\nu \ell$ where $\nu$ is the effective quantum number, 
converging on to the target threshold $S_iL_I$.

The first term on the right in Eq.
(1) represents both bound states and free (continuum) states of the
\eion system. If all channels are closed then the state is bound an
represented by $\Psi_B$; asymptotically, all channel functions are 
expnentially decaying.
On the other hand, a continuum state corresponds
to some channels open and some closed, and is referred to as $\Psi_F$.
Asymptotically, the open channel functions are oscillating, as for a
free electron. The transition matrix element for photoionization is 
\begin{equation}
<\Psi_B || {\bf D}|| \Psi_{F}>,
\end{equation}
where ${\bf D} = \sum_i{r_i}$ is the dipole operator and the sum is over 
the number of electrons; $\Psi_B$ and $\Psi_{F}$ are the bound and 
continuum wave functions. The transition matrix element can be reduced 
to generalized line strength as
\begin{equation}
{\bf S}= |<\Psi_j||{\bf D}_L||\Psi_i>|^2 =
 \left|\left\langle{\mit\psi}_f
 \vert\sum_{j=1}^{N+1} r_j\vert
 {\mit\psi}_i\right\rangle\right|^2 \label{eq:SLe},
\end{equation}
where $\mit\Psi_i$ and $\mit\Psi_f$ are the initial and final state
wave functions. The photoionization cross section ($\sigma_{PI}$) is
proportional to the generalized line strength as,
\begin{equation}
\sigma_{PI} = {4\pi^2 \over 3c}{1\over g_i}\omega{\bf S},
\end{equation}
where $g$ is the statistical weight factor of the bound state and
$\omega$ is the incident photon energy (Ry).

\section{Computations}

BPRM photoionization cross sections of \fexvii were computed in the 60CC 
expansion over the core ion \fexviii, and free-electron wave functions 
with partial waves up to $l\leq$ 12. The one-electron basis set of 
orbitals used to construct a configuration-interaction representation of 
the eigenfunctions of \fexviii were obtained from the code SUPERSTRUCTURE 
\cite{ss}, which employs a Thomas-Fermi-Dirac-Amaldi central field 
potential. The choice of the one-electron orbital basis set is not unique; 
Hartree-Fock orbitals of Slater type orbitals are also used in R-matrix 
calculations \cite{br75,burke,bb93}. At sufficiently high densities, viz. 
$N_e > 10^{22}$ cc, the electronic orbitals may themselves be altered by 
plasma effects such as Debye screening. However, we do not consider those 
effects in OP calculations which entail the isolated atom approximation 
and the so called "chemical picture" \cite{mhd, symp} (discussed later).

Table~1 gives the set of 11 configurations with filled 1s$^2$ shell that 
was optimized to obtain \fexviii core wave functions \cite{znp}. It lists 
the 60 levels included in the 60CC expansion. For a small number of levels 
the available energy level data \cite{sugar} compiled by the National 
Institute for Standards and Technology (NIST), is found to be in very good 
agreement with theoretical calculations \cite{znp}. One of the main points 
to note, and which is of considerable significance in the {\it energy 
distribution of resonances}, is the clustering of 60 fine structure levels 
in the $n$ = 2 and $n$ = 3 complexes, and the large energy gap of $\sim$50 
Ry between them.
\begin{table}
\caption{Energy levels of the target ion \fexviii included in the 
eigenfunction expansion of \fexvii. Note the large energy gap of $\sim$47 
Ry between the \en = 2 and \en = 3 complexes. The target was optimized 
with a set of 11 configurations with open L-shell but closed K-shell: 
2s$^2$2p$^5$(1), 2s2p$^6$(2),
2s$^2$2p$^4$3s(3), 2s$^2$2p$^4$3p(4), 2s$^2$2p$^4$3d(5), 2s2p$^5$3s(6),
2s2p$^5$3p(7), 2s2p$^5$3d(8), 2p$^6$3s(9), 2p$^6$3p(10), 2p$^6$3d(11).
\label{table1}}
\scriptsize
\begin{tabular}{rcccrccccc}
\noalign{\smallskip}
\hline
\noalign{\smallskip}
 i & Config- &Term  &2$J$&  $E$(Ry) &
 i & Config- &Term  &2$J$&  $E$(Ry) \\
 & uration & & & Present &  & uration & & & Present \\
\noalign{\smallskip}
\hline
\noalign{\smallskip}
\multicolumn{5}{c}{n=2~states} &
30 & $2s^22p^43p$   &$^2P^o$ & 1  &   61.899   \\
 1 &   $2s^22p^5$   &$^2P^o$ & 3  &   0.00000 &
31 &   $2s^22p^43d$ &$^4D$   & 5  &   62.299  \\
 2 &   $2s^22p^5$   &$^2P^o$ & 1  &  0.93477 &   
32 &  $2s^22p^43d$  &$^4D$   & 7  &   62.311  \\ 
 3 &   $2s  2p^6$   &$^2S$   & 1  &  9.70228 &
33 &  $2s^22p^43d$  &$^4D$   & 1  &  62.906  \\
\multicolumn{5}{c}{n=3~states} &
34 &  $2s^22p^43d$  &$^4D$ & 3  &   63.050  \\
 4 &   $2s^22p^43s$ &$^4P$ & 5  &   56.690   & 
35 &   $2s^22p^43p$ &$^2P^o$ & 3  &   62.461  \\
 5 &   $2s^22p^43s$ &$^2P$ & 3  &   56.936   &
36 &   $2s^22p^43d$ &$^4F$ & 9  &   62.535   \\
 6 &   $2s^22p^43s$ &$^4P$ & 1  &   57.502   &
37 & $2s^22p^43d$   &$^2F$ & 7  &   62.629  \\
 7 &   $2s^22p^43s$ &$^4P$ & 3  &   57.572   &
38 &   $2s^22p^43p$ &$^2P^o$ & 1  &  62.686  \\
 8 &   $2s^22p^43s$ &$^2P$ & 1  &   57.798  &
39 &   $2s^22p^43d$ &$^4P$ & 1  &   62.496  \\
 9 &   $2s^22p^43s$ &$^2D$ & 5  &   58.000  &
40 & $2s^22p^43d$   &$^4P$ & 3  &   62.625      \\
10 &   $2s^22p^43s$ &$^2D$ & 3  &   58.355   &
41 & $2s^22p^43d$   &$^4F$ & 5  &   62.985   \\
11 &   $2s^22p^43p$ &$^4P^o$ & 3  &  59.209   &
42 & $2s^22p^43d$   &$^2P$ & 1  &   63.123   \\
12 &   $2s^22p^43p$ &$^4P^o$ & 5  &  59.238   &
43 & $2s^22p^43d$   &$^4F$ & 3  &   63.156    \\
13 &  $2s^22p^43p$  &$^4P^o$ & 1 &   59.478   &
44 & $2s^22p^43d$   &$^2F$ & 5  &   62.698      \\
14 & $2s^22p^43p$   &$^4D^o$ & 7  &   59.525   &
45 & $2s^22p^43d$   &$^4F$ & 7  &   63.271    \\
15 & $2s^22p^43p$   &$^2D^o$ & 5  &   59.542   &
46 & $2s^22p^43d$   &$^2D$ & 3  &   63.302    \\
16 &   $2s^22p^43s$ &$^2S$ & 1  &   59.916      &
47 & $2s^22p^43d$   &$^4P$ & 5  &   62.911      \\
17 &   $2s^22p^43p$ &$^2P^o$ & 1  &   59.982   &
48 & $2s^22p^43d$   &$^2P$ & 3  &   63.308      \\
18 &  $2s^22p^43p$  &$^4D^o$ & 3  &   60.005   &
49 & $2s^22p^43d$   &$^2D$ & 5  &   63.390      \\
19 &  $2s^22p^43p$  &$^4D^o$ & 1  &   60.012   &
50 & $2s^22p^43d$   &$^2G$ & 7  &   63.945   \\
20 & $2s^22p^43p$   &$^2D^o$ & 3  &   60.147   &
51 & $2s^22p^43d$   &$^2G$ & 9  &   63.981    \\
21 & $2s^22p^43p$   &$^4D^o$ & 5  &   60.281   &
52 & $2s^22p^43d$   &$^2S$ & 1  &   63.919      \\
22 & $2s^22p^43p$   &$^2P^o$ & 3  &   60.320   &
53 & $2s^22p^43d$   &$^2F$ & 5  &   64.200    \\
23 &  $2s^22p^43p$  &$^2S^o$ & 1  &   60.465   &
54 & $2s^22p^43d$   &$^2F$ & 7  &   64.301    \\
24 & $2s^22p^43p$   &$^4S^o$ & 3  &   60.510   &  
55 & $2s^22p^43d$   &$^2P$ & 3  &   64.138      \\
25 & $2s^22p^43p$   &$^2F^o$ & 5  &   60.851   &
56 & $2s^22p^43d$   &$^2D$ & 5  &    64.160      \\
26 & $2s^22p^43p$   &$^2F^o$ & 7  &   61.028   &
57 & $2s^22p^43d$   &$^2D$ & 3  &   64.391      \\
27 & $2s^22p^43p$   &$^2D^o$ & 3  &   61.165   & 
58 & $2s^22p^43d$   &$^2P$ & 1  &    64.464      \\
28 & $2s^22p^43p$   &$^2D^o$ & 5  &   61.272   & 
59 & $2s^22p^43d$   &$^2D$ & 5  &   65.305      \\
29 & $2s^22p^43p$   &$^2P^o$ & 3  &   61.761   & 
60 & $2s^22p^43d$   &$^2D$ & 3  &   65.468      \\
\noalign{\smallskip}
\hline
\end{tabular}
\end{table}
Although the target energies are quite accurate, some further
improvement in the positions of resonances is also achieved  by
replacing the \fexviii level energies with observed ones, wherever
available \cite{sugar}, during diagonalization of the (N+1)-electron 
Hamiltonian of the (e~+~\fexviii) system.

The second sum in the wavefunction expansion given in Eq. (1) is the 
bound-channel term consisting of selected electronic configurations
for the electron-ion system. We include 27 configurations of 
({\it N}+1)-electron bound channels of \fexvii,  specified by a range 
of mininum and maximum occupancies (listed within parentheses after 
the orbitals) as: $2s(0-2)$, $2p(3-6)$, $3s(0-2)$, $3p(0-2)$ and 
$3d(0-2)$. All $SLJ\pi$ symmetries of the electron-ion system formed 
from the target states coupled with an interacting electron with 
continnum partial waves 0 $\leq l \leq$ 12 are considered.

The ab initio bound state energies of the electron-ion system computed 
by the R-matrix codes in intermediate $SLJ$ coupling are not 
spectroscopically identified {\it a priori}. For complex ions it is a 
highly non-trivial task to assign $LS$ term and $SLJ$ level designations. 
It is particularly difficult for fine structure levels computed in BPRM 
calculations owing to near-degeneracy of levels in high-$Z$ or high-$z$
ions, particularly for high-\en and $l$. The code PRCBPD \cite{np00} is 
employed for level identification, using the information computed by the 
BPRM codes, as explained in the next section. In addition, the BPRM 
energies are cross checked against observations and SUPERSTRUCTURE 
whereever possible. However, we still note that this is one of the most 
laborious tasks requiring some judgement in the final assignments for 
levels highly mixed by configuration interaction.

The present calculations cover a large energy range and variations in 
the distribution of resonance complexes. In the near-threshold region, 
below $n$ = 2 levels, the $\sigma_{PI}$ were resolved on a fine energy 
mesh, with 4000 energies upto 0.4 Ry above the ionization threshold. 
However, such fine resolution is computationally prohibitive for all 
symmetries and levels over the entire range of $\sim$65 Ry where 
resonances due to L-shell excitation occur. There are energy regions 
where the resonances are sparse, as opposed to regions where they are 
densely clustered. A number of energy meshes are used to ensure that 
the overall, as well as the detailed, contribution of resonances to 
$\sigma_{PI}$ is taken into account. At photoelectron energies above 
all 60 target thresholds, the $\sigma_{PI}$ are extrapolated as in 
\cite{np94}.

\section{Results and Discussion}

The results of \fexvii are divided into a few subsections below.

\subsection{Fine Structure Levels and Oscillator Strengths of \fexvii}

Present BPRM calculations for fine structure bound levels are intended 
to form a complete set for most practical applications. As explained in 
an earlier paper \cite{n08}, the high-lying excited core states do not 
form bound states of the electron-ion system, and corresponding channels 
have insignificant effect on the bound state energies. This is 
particularly true of multiply ionized ions, where there are large gaps 
between the ground complex and the next excited complex of levels; the 
energy separation increases as $z^^2$. The core ion \fexviii has an 
energy gap of $\sim$47 Ry between the $n$=2 and $n$=3 levels (Table 1). 
Although the present study includes both the n=2 and 3 complexes, all 
bound levels of \fexvii have the \fexviii parent level as the ground 
level or another $n$=2 excited level, i.e.  $2s^2p^5 \ ^2P^o_{1/2,3/2}, 
\ 2s2p^6 \ ^2S_{1/2}$.

R-matrix calculations for bound energy levels entail a `search' for 
zeroes of the electron-ion Hamiltonian \cite{mjs87,ip}, where the 
eigenvalues of the (N+1)-electron system occur as in Eq. (4). A total 
of 454 eigenenergies of \fexvii were found, subject to the choice of 
$n\leq$ 10, 0 $\leq l \leq$ 9, and 0 $\leq J\leq$ 8 of even and odd 
parities. These levels have been identified spectroscopically using a 
numerical procedure implemented in the code PRCBPID \cite{np00}. It is 
based on (i)  detailed analysis of quantum defects along Rydberg series 
of levels, (ii) parentage of the core ion states, (iii) fractional or 
percentage contribution of closed channels which translates into 
configuration interaction of corresponding bound electron-ion 
configurations, and (iv) angular momentum algebra. The level and 
energies are similar to the 3CC case carried \cite{znp}, with some 
differences because of mixed levels and near-degenerate quantum defects.

For relatively few levels the present calculated  energies of \fexvii 
are compared with the observed values available in the NIST compilation
\cite{sugar} in Table~2. The overall agreement is very good, $\sim$0.1\% 
or better, for most levels including ones that are highly excited. The
number next to the $J$-value in Tabel~2 (Col. 3) is the relative position 
of the correspoding calculated energy in its own symmetry $J\pi$. As an 
example we discuss levels of a particular summetry $SLJ:^1P^o_1$. In the 
present work we obtain 35 $^1P^o_1$ levels, as opposed to 11 given in 
the NIST tables; they are compared in Table~2. Of those 11 energies, 5 
energies agree to $\sim$0.1\% or better. That includes highly excited 
levels such as $2s^22p^64p~^1P^o_1$, listed as the 19$^{th}$ computed 
level (Col. 3). For two other levels, $2s^22p^56d~^1P^o_1$ and 
$2s^22p^55d~^1P^o_1$, the energies agree to $\sim$ 0.2\% and 0.3\% 
respectively. However, the energies of the remaining three levels differ 
by 1-4\%. The largest discrepancy is for the level $2s^22p^55s~^1P^o_1$. 
The complete set of 454 energies with spectroscopic designations is 
available electronically (at the NORAD atomic data website given at the 
end of the paper).
\begin{table}
\caption{Comparison of calculated energies, $E_c$, of Fe~XVII with the
measured values, $E_o$ \cite{sugar}. $i_J$ indicates position of the
calculated level for symmetry $J$. An asterisk next to a level indicates 
incomplete set of observed levels for the state.}
\scriptsize
\begin{tabular}{llcll}
\noalign{\smallskip}
\hline
            \noalign{\smallskip}
\multicolumn{1}{c}{Conf} & \multicolumn{1}{c}{Term} & $J:i_J$&
\multicolumn{1}{c}{$E_o$(Ry)} & \multicolumn{1}{c}{$E_c$(Ry)} \\
            \noalign{\smallskip} 
\hline
            \noalign{\smallskip}
$2s^22p^6   $&         $^1S  $&   0.0 :1 &   9.2760E+01&   9.2925E+01\\ 
$2s^22p^53s $&         $^3P^o$&   2.0 :1 &   3.9463E+01&   3.9503E+01\\
$2s^22p^53s $&         $^3P^o$&   1.0 :1 &   3.9323E+01&   3.9367E+01\\
$2s^22p^53s $&         $^3P^o$&   0.0 :1 &   3.8533E+01&   3.8560E+01\\
$2s^22p^53s $&         $^1P^o$&   1.0 :2 &   3.8446E+01&   3.8469E+01\\
$2s^22p^53p $&         $^3S  $&   1.0 :1 &   3.7238E+01&   3.7284E+01\\
$2s^22p^53p $&         $^3D  $&   3.0 :1 &   3.6863E+01&   3.6902E+01\\
$2s^22p^53p $&         $^3D  $&   2.0 :1 &   3.6981E+01&   3.7027E+01\\
$2s^22p^53p $&         $^3D  $&   1.0 :3 &   3.6093E+01&   3.6114E+01\\
$2s^22p^53p $&         $^1P  $&   1.0 :2 &   3.6780E+01&   3.6826E+01\\
$2s^22p^53p $&         $^3P  $&   2.0 :2 &   3.6646E+01&   3.6688E+01\\
$2s^22p^53p $&         $^3P  $&   1.0 :4 &   3.5854E+01&   3.5880E+01\\
$2s^22p^53p $&         $^3P  $&   0.0 :2 &   3.6244E+01&   3.6274E+01\\
$2s^22p^53p $&         $^1D  $&   2.0 :3 &   3.5826E+01&   3.5843E+01\\
$2s^22p^53p $&         $^1S  $&   0.0 :3 &   3.4871E+01&   3.4828E+01\\
$2s^22p^53d $&         $^3P^o$&   2.0 :2 &   3.3662E+01&   3.3669E+01\\
$2s^22p^53d $&         $^3P^o$&   1.0 :3 &   3.3778E+01&   3.3813E+01\\
$2s^22p^53d $&         $^3P^o$&   0.0 :2 &   3.3862E+01&   3.3895E+01\\
$2s^22p^53d $&         $^3F^o$&   4.0 :1 &   3.3656E+01&   3.3651E+01\\
$2s^22p^53d $&         $^3F^o$&   3.0 :1 &   3.3599E+01&   3.3612E+01\\
$2s^22p^53d $&         $^3F^o$&   2.0 :4 &   3.2672E+01&   3.2670E+01\\
$2s^22p^53d $&         $^1D^o$&   2.0 :3 &   3.3472E+01&   3.3494E+01\\
$2s^22p^53d $&         $^3D^o$&   3.0 :2 &   3.3393E+01&   3.3400E+01\\
$2s^22p^53d $&         $^3D^o$&   2.0 :5 &   3.2598E+01&   3.2601E+01\\
$2s^22p^53d $&         $^3D^o$&   1.0 :4 &   3.3052E+01&   3.3053E+01\\
$2s^22p^53d $&         $^1F^o$&   3.0 :3 &   3.2563E+01&   3.2565E+01\\
$2s^22p^53d $&         $^1P^o$&   1.0 :5 &   3.2070E+01&   3.2049E+01\\
$2s2p^63p  $&         $^3P^o$&   1.0*:6 &   2.7159E+01&   2.7122E+01\\
$2s2p^63p  $&         $^1P^o$&   1.0 :7 &   2.6836E+01&   2.6809E+01\\
$2s^22p^54s $&         $^3P^o$&   1.0*:8 &   2.0899E+01&   2.0630E+01\\
$2s^22p^54s $&         $^1P^o$&   1.0 :9 &   2.0014E+01&   2.0557E+01\\
$2s^22p^54d $&         $^3P^o$&   1.0*:10&   1.8802E+01&   1.8750E+01\\
$2s^22p^54d $&         $^3D^o$&   1.0*:11&   1.8455E+01&   1.8427E+01\\
$2s^22p^54d $&         $^1P^o$&   1.0 :12&   1.7590E+01&   1.7571E+01\\
$2s^22p^55s $&         $^3P^o$&   1.0*:13&   1.2960E+01&   1.2739E+01\\
$2s^22p^55s $&         $^1P^o$&   1.0 :14&   1.2022E+01&   1.2516E+01\\
$2s^22p^55d $&         $^3P^o$&   1.0*:15&   1.2022E+01&   1.1912E+01\\
$2s^22p^55d $&         $^3D^o$&   1.0*:16&   1.1776E+01&   1.1749E+01\\
$2s^22p^55d $&         $^1P^o$&   1.0 :17&   1.0910E+01&   1.0873E+01\\
$2s2p^64p  $&         $^3P^o$&   1.0*:18&   1.0236E+01&   1.0197E+01\\
$2s2p^64p  $&         $^1P^o$&   1.0 :19&   1.0090E+01&   1.0080E+01\\
$2s^22p^56s $&         $^3P^o$&   1.0*:20&   8.7776E+00&   8.7515E+00\\
$2s^22p^56d $&         $^3P^o$&   1.0*:22&   8.1488E+00&   8.1405E+00\\
$2s^22p^56d $&         $^1P^o$&   1.0*:24&   7.2558E+00&   7.2410E+00\\
$2s^22p^57s $&         $^3P^o$&   1.0*:25&   6.3810E+00&   6.3535E+00\\
$2s^22p^57d $&         $^3P^o$&   1.0*:26&   5.9709E+00&   6.0260E+00\\
$2s^22p^57d $&         $^1P^o$&   1.0*:29&   5.0232E+00&   5.0588E+00\\
$2s^22p^58d $&         $^3P^o$&   1.0*:32&   4.4582E+00&   4.5627E+00\\
$2s^22p^58d $&         $^1P^o$&   1.0*:35&   3.6016E+00&   3.6512E+00\\
\hline
\end{tabular}
\end{table}

We also report that the oscillator strengths for electric dipole transitions
among all the bound levels are obtained and are available electronically. 
The earlier set corresponds to 3-CC calculations \cite{n03}.

\subsection{Photoionization Cross Sections of \fexvii}

Photoionization cross sections of all 454 bound fine structure levels
of \fexvii are computed and analyzed in a variety of ways, in particular
the distribution of resonances that lie in the large energy gap between
the $n$ = 2 and $n$ = 3 complexes. This is a comprehensive set of 
cross sections computed using the BPRM method, as required for 
applications such as plasma opacities, synthetic spectral models, and 
spectral diagnostics of X-ray absorption lines observed from astronomical 
objects. The BPRM calculations yield about two and a half times as many
levels as the previous OP calculations in LS coupling \cite{top}, which 
resulted in 181 LS bound states with $n(SL) \leq 10$. Some important 
charateristic features of the BPRM cross sections are illustrated and 
discussed below.

\subsubsection{Photoionization of the Ground State}

First we consider the ground state of \fexvii.  Fig.~1 presents 
photoionization cross sections $\sigma_{PI}$ of the ground state 
$2s^22p^6 ~^1S$. While panel (a) presents the cross section in LS coupling 
under the Opacity Project (M. P. Scott, see TOPbase data \cite{top}), 
panels (b,c) present $\sigma_{PI}$ from the relativistic BPRM 
calculations; (b) presents the total cross section, and (c) presents 
partial cross section for leaving the core ion \fexviii in the ground 
state following photoionization. The earlier total $\sigma_{PI}$ from 
the two-state R-matrix OP calculation, and the present BPRM using a 
60-level expansion are of similar magnitude for the background. But due 
to a much larger number of fine structure channels in the BPRM 
calculation, panel (b) shows many more resonances within the $n$=2 
complex (as also found in the earlier 3CC calculation). Also, a much 
finer energy mesh in the present work has resolved the resonances more 
completely. Resolution of resonances in the low energy region, especially 
at and near threshold is crucial for calculating recombination and 
photoionization rates. The ground state $\sigma_{PI}$ at high energies 
decreases slowly, showing an insignificant effect due to highly excited 
$n$=3 core states in the 60CC calculation, except for introducing small 
and weak resonance structures.
\begin{figure}
\resizebox{90mm}{!}
{ \includegraphics{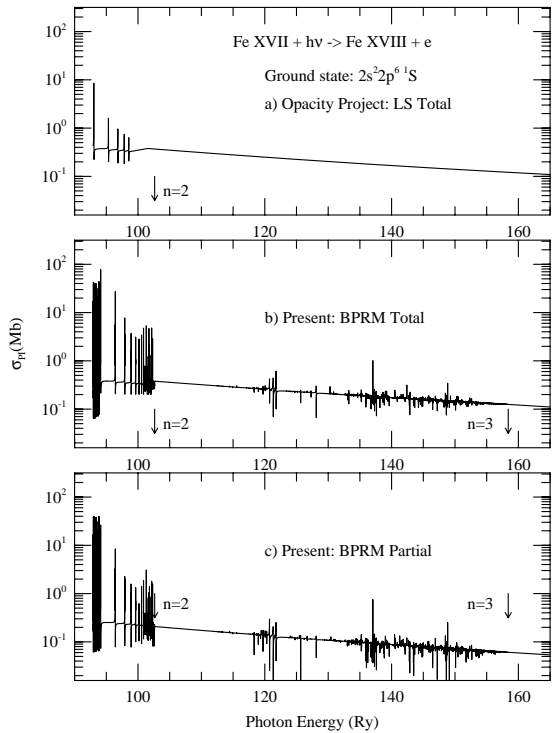}}
\caption{Photoionization cross sections, $\sigma_{PI}$, of the ground
level $1s^22s^22p^6(^1S_0)$ of \fexvii: (a) total $\sigma_{PI}$ in LS 
coupling from the Opacity Project database TOPbase \cite{top}, (b) 
present total $\sigma_{PI}$ in the relativistic BPRM formulation, (c) 
present partial $\sigma_{PI}$ with residual ion \fexviii in the ground 
state. Present calculations including relativistic fine structure
provide more accurate energies and more highly resolved resonances. The 
background for the partial cross sections (c) is lower that the total (b)
because the channels for leaving the core ion in various excited levels 
are excluded.}
\end{figure}

The partial cross sections in panel (c) corresponds to photoionization
of \fexvii leaving the ionized core \fexviii only in the ground state. 
The background is lower than that of the total $\sigma_{PI}$ in (b) 
because there are no additional contributions from photoionization 
channels into excited levels. However the resonance features are the 
same for the total and the partial cross sections until the first excited 
state $2s2p^6 ~^1S_0$ of \fexviii.

\subsubsection{Photoionization of Excited Levels}

Although the background ground state cross sections remain about the
same between the 3CC and the 60CC calculations, the excited states are 
considerably more affected. One of the main results of the present BPRM 
calculations with a large wavefunction expansion is that photon 
absorptions and core excitations to the $n$=3 levels produce extensive 
resonance structures, in contrast to earlier calculations. This 
contradicts the assumption that channels of highly excited core states, 
especially when an energy gap like in the present case exists, might be 
too weakly interacting to produce any significant effect.

  \begin{figure}
   \centering
\resizebox{90mm}{!}{
  \includegraphics{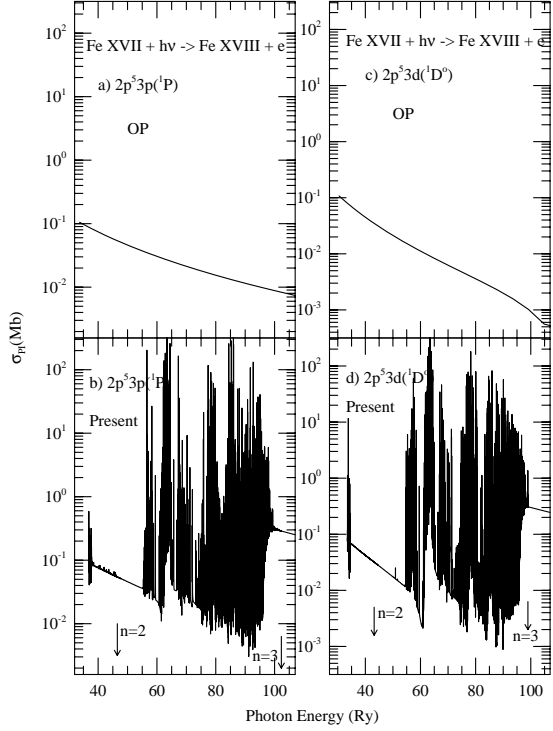}}
\caption{Comparison of present photoionization cross sections 
$\sigma_{PI}$ with those from the OP (see \cite{top}) for excited 
singlet levels (no fine structure): (a,b)  $2p^53p^1P$, and (c,d) 
$2p^53d(^1D^o)$. In contrast to $\sigma_{PI}$ from OP, the present 
results demonstrate that without inclusion of $n$=3 core levels of 
\fexviii the cross sections are considerably underestimated throughout 
most of the energy region of interest in practical applications. 
Resonances are included as lines in the OP work \cite{op}, as in other 
opacities calculations \cite{opal}.}
              \label{FigGam}%
   \end{figure}
Fig.~2 presents total photoionization cross sections of excited levels 
$2s^22p^53p(^1P^o_1)$ (a,b), and $2s^22p^53d(^1D^o_2)$ (c,d). These two 
levels are chosen because they have a single fine structure component, so 
their cross sections can be directly compared with the earlier OP 
results for the LS terms $2s^22p^53p(^1P^o)$ and $2s^22p^53d(^1D^o)$. The 
top panels (a,c) of Fig. 2 present the OP  results, and the bottom panels 
(b,d) from the present work. Both of these excited levels show resonances 
due to core excitations to the $n = 2$ and \en = 3 levels that are not 
present in the OP results. The features reveal how crucially important 
the set of resoances belonging to $n$=3 core thresholds are, compared to 
those of $n$ = 2. The resonance peaks are orders of magnitude higher than 
those from the $n$=2 series. There is only a relatively small gap with 
smooth background without resonances, between the highest threshold of 
the $n$=2 levels and the appearance of lowest of the $n$=3 complexes of 
resonances. While the background is enhanced considerably in the BPRM 
results by strong resonances, the featureless and slowly decreasing 
cross sections from the OP miss out all those features and vastly 
underestimate $\sigma_{PI}$.

Resonances in photoionization cross sections correspond to Rydberg series 
of autoionizing levels at energies $E_{res} = E_c\nu l$, where $E_c$ is 
an excited core threshold and $\nu_l$ are the efeective quantum number 
and angulur momentun of the interacting electron. The Rydberg resonances
are narrow and lie below the threshold $E_c$, approximately at energies
given by the simple expression 
\begin{equation}
E_{res} = E_c -z^2/\nu_l^2.
\end{equation}
However, strong PEC resonances manfiest themselves in $\sigma_{PI}$ of 
excited bound levels of the electron-ion system \cite{ys87,np91}. The 
PEC resonances are wide and occur at energies where the ground state of 
the core ion undergoes a strong dipole allowed transition. Among the 60 
target or core levels for \fexviii~ there are 30 such transitions as 
given in Table~3. Hence, in photoionization cross sections of \fexvii 
there are 30 possible PEC resonances, most with overlapping profiles.
\begin{table}
\scriptsize
\caption{Dipole allowed and intercombination E1 transitions from the 
ground level $2s^22p^5(^2P^o_{3/2})$ to excited states of the core ion 
\fexviii, and corresponding oscillator strengths ($f$). These transitions 
introduce the PEC resonances in photoionization cross sections 
of \fexvii. A number 'a(n)' means a$\times$10$^n$. The level indices 
correspond to those of Table 1.
}
\begin{tabular}{lll}
\hline
\noalign{\smallskip}
\multicolumn{1}{c}{Levels} & \multicolumn{1}{c}{Transition} &
f(PEC)\\
\noalign{\smallskip}
\hline 
\noalign{\smallskip}
1-3&$2s^22p^5(^2P^o_{3/2})-2s2p^6(^2S_{1/2})$ &  5.85E-02\\
1-4&$2s^22p^5(^2P^o_{3/2})-2s^22p^43s(^4P_{5/2})$ & 3.57E-03\\
1-5&$2s^22p^5(^2P^o_{3/2})-2s^22p^43s(^2P_{3/2})$ & 4.53E-02\\
1-6&$2s^22p^5(^2P^o_{3/2})-2s^22p^43s(^2P_{1/2})$ & 1.54E-02\\
1-7&$2s^22p^5(^2P^o_{3/2})-2s^22p^43s(^4P_{3/2})$ & 2.12E-02\\
1-8&$2s^22p^5(^2P^o_{3/2})-2s^22p^43s(^4P_{1/2})$ & 1.51E-03\\
1-9&$2s^22p^5(^2P^o_{3/2})-2s^22p^43s(^2D_{5/2})$ & 3.43E-02\\
1-10&$2s^22p^5(^2P^o_{3/2})-2s^22p^43s(^2D_{3/2})$ & 3.87E-04\\
1-20&$2s^22p^5(^2P^o_{3/2})-2s^22p^43s(^2S_{1/2})$ & 2.86E-03\\
1-31&$2s^22p^5(^2P^o_{3/2})-2s^22p^43d(^4D_{5/2})$ & 2.84E-07\\
1-33&$2s^22p^5(^2P^o_{3/2})-2s^22p^43d(^4D_{3/2})$ & 1.43E-04\\
1-34&$2s^22p^5(^2P^o_{3/2})-2s^22p^43d(^4D_{1/2})$ & 3.88E-02 \\
1-37&$2s^22p^5(^2P^o_{3/2})-2s^22p^43d(^4P_{1/2})$ & 8.65E-05\\
1-39&$2s^22p^5(^2P^o_{3/2})-2s^22p^43d(^4P_{3/2})$ & 1.31E-01\\
1-41&$2s^22p^5(^2P^o_{3/2})-2s^22p^43d(^4P_{5/2})$ & 3.46E-01\\
1-42&$2s^22p^5(^2P^o_{3/2})-2s^22p^43d(^2P_{1/2})$ & 4.10E-03\\
1-43&$2s^22p^5(^2P^o_{3/2})-2s^22p^43d(^2D_{3/2})$ & 3.58E-02\\
1-44&$2s^22p^5(^2P^o_{3/2})-2s^22p^43d(^2F_{5/2})$ & 3.92E-04\\
1-46&$2s^22p^5(^2P^o_{3/2})-2s^22p^43d(^4F_{3/2})$ & 9.34E-03\\
1-47&$2s^22p^5(^2P^o_{3/2})-2s^22p^43d(^4P_{5/2})$ & 1.38E-02\\
1-48&$2s^22p^5(^2P^o_{3/2})-2s^22p^43d(^2P_{3/2})$ & 1.46E-02\\
1-49&$2s^22p^5(^2P^o_{3/2})-2s^22p^43d(^4F_{5/2})$ & 2.39E-01\\
1-52&$2s^22p^5(^2P^o_{3/2})-2s^22p^43d(^2F_{5/2})$ & 3.37E-02\\
1-53&$2s^22p^5(^2P^o_{3/2})-2s^22p^43d(^2S_{1/2})$ & 1.66E-01\\
1-55&$2s^22p^5(^2P^o_{3/2})-2s^22p^43d(^2P_{3/2})$ & 3.84E-01\\
1-56&$2s^22p^5(^2P^o_{3/2})-2s^22p^43d(^2D_{5/2})$ & 5.08E-01\\
1-57&$2s^22p^5(^2P^o_{3/2})-2s^22p^43d(^2D_{3/2})$ & 6.81E-02\\
1-58&$2s^22p^5(^2P^o_{3/2})-2s^22p^43d(^2P_{1/2})$ & 4.08E-02\\
1-59&$2s^22p^5(^2P^o_{3/2})-2s^22p^43d(^2D_{5/2})$ & 5.24E-02\\
1-60&$2s^22p^5(^2P^o_{3/2})-2s^22p^43d(^2D_{3/2})$ & 3.72E-03\\
\hline
\end{tabular}
\end{table}

Fig.~3 presents photoionization cross sections of three highly excited
levels of \fexvii and demonstrably large PEC features, $2s^22p^5nf(^3D_1)$ 
with $nf=5f,7f,9f$. Beyond the resonances due to $n$ = 2 core levels, the 
background decreases smoothly for the three levels. However, as the 
resonances due to \en = 3 core levels appear, the backgound rises and the 
PEC resonances manifest themselves as prominently high and wide structures.
These enhancements are related to the radiative decay rates of the
dipole allowed levels. Table~3 shows most of the decay rates from \en = 3
levels are much larger than those from the \en = 2 levels, 
by one or two orders of magnitude, and result in stronger resonances. 
With hundreds of bound levels computed in the present work, the cross
sections in the high energy 
region from about 57 Ry to 65 Ry is greatly enhanced, mainly by complexes 
of PEC resonances corresponding to core excitations via dipole transitions 
listed in Table~3.
   \begin{figure}
   \centering
\resizebox{90mm}{!}
{\includegraphics{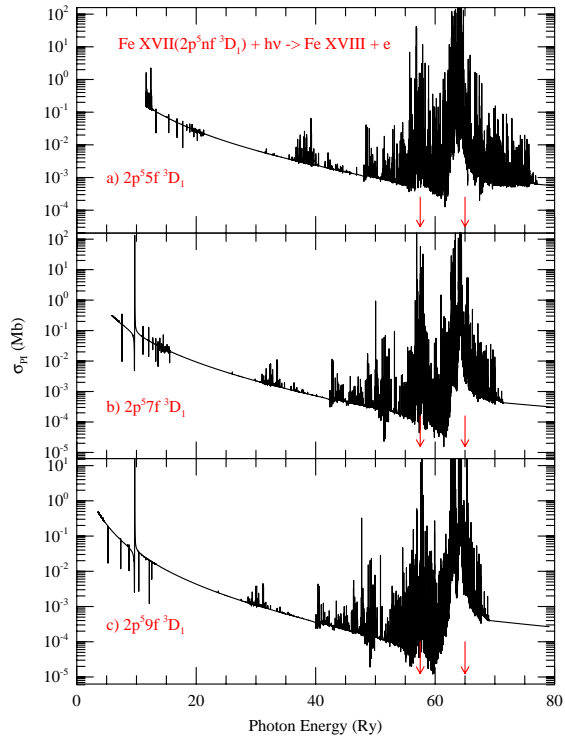}}
   \caption{Total photoionization cross section $\sigma_{PI}$ of highly
excited levels of \fexvii: $2s^22p^5nf(^3D_1)$ with $nf=5f,7f,9f$. The
cross sections decrease monotonically at lower energies, but rise
considerably while approaching the PEC resonances. The figure shows that
the cross sections would be enhanced all throughout the energy range,
compared to currently available near-hydrogenic cross sections that
decrease much faster as $\sigma_{PI} \sim E^{-3}$. The largest PEC
resonances are marked by arrows. Although only two such PECs are pointed
out, the resonance structures and the resulting enhancement represent
the combined effect of all PEC resonances in Table~3. It may be also be
noted that the PEC positions remain the same for all Rydberg levels.}
              \label{FigGam}%
   \end{figure}

Of particular importance is the characterstic shape of PEC resonances.
The resonances encompass an energy range of $\sim$10 Ry or well over
100 eV. The shape is determined by channel coupling effects. In the case 
of the simple limit of an isolated resonance it is like the typical Fano 
profile. However, the PECs are generally affected by strong coupling among
many target levels and channels, with a large number of superimposed
non-PEC resonances converging on to the target levels. The distribution
of the continuum or the differential oscillator strength would not be
generally reproduced by an isolated resonance approximation, and requires 
a coupled channel calculation.

\subsection{Resonance oscillator strengths}

A fundamental approximation made in existing opacities calculations is 
to treat autoionizing resonances as lines. In other words, inner-shell 
excitations leading to bound-free autoionizing states are treated as 
bound-bound transitions. The final state may be further coupled to a 
feature-less continuum perturbatively to obtain autoionization widths 
at a single energy associated with the bound-bound transition. This 
indepndent resonance approximation neglects the intricate coupling 
effects that are otherwise included via the close coupling method. Since 
most of the contribution to opacities originates from inner-shell 
excitations, with final levels as autoionizing states, their impact on 
opacities bears closer inspection and is briefly discussed below.

   \begin{figure}
   \centering
\resizebox{90mm}{!}
{ \includegraphics{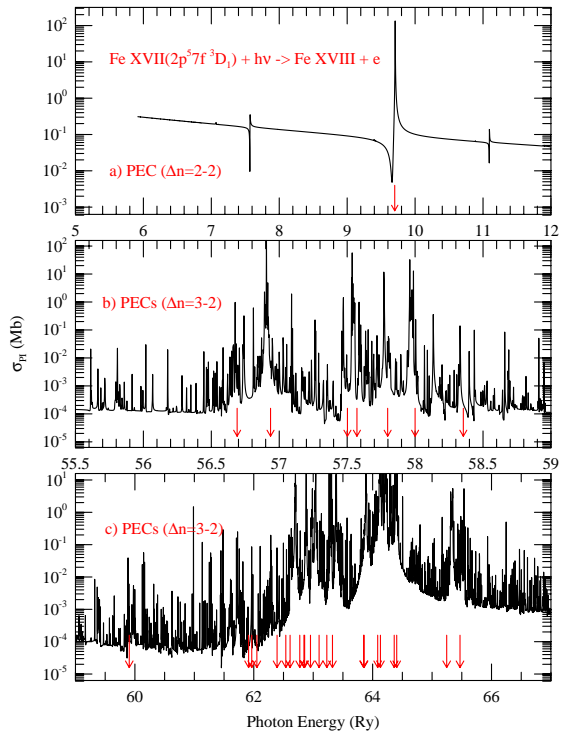}}
   \caption{L-shell PEC resonances in photoionization of \fexvii. The 
resonances cover all 60 ionization thresholds included in
the CC calculation, including all PEC resonances corresponding to E1
transitions (dipole allowed and intercombination) in Table~3.}
              \label{FigGam}%
   \end{figure}
In Fig.~4 we examine in some detail one of the cross sections shown in 
Fig.~3, corresponding to $2s^22p^5 \ 7f(^3D_1)$. The cross sections are 
delineated at 24,653 energies; the energy intervals are chosen so as to 
resolve resonance profiles in so far as practical. Postions of all of the 
PEC resonances due to transitions in Table~3 are marked in Fig.~4 as 
arrows. In addition, there are a large number of non-PEC Rydberg resonances 
converging on to the other excited levels. Whereas the few $\Delta n(2-2)$ 
resonances in Fig. 4a are inconsequential, the myriad $\Delta n(3-2)$ PECs 
in Fig. 4b and 4c dominate the distribution of oscillator strength over a 
large region $\sim$56-66 Ryd, or $>$100 eV, due to the L-shell excitation 
array $2p \rightarrow 3d$. The cumulative resonance (or bound-free 
continuum) oscillator strength corresponding to Fig.~4 is related to the 
photoionization cross section (viz. \cite{nzp}) as
\begin{eqnarray}
f_r [(^3D_1 \longrightarrow \epsilon SLJ: ^3(P,D,F) \ (J = 0,1,2)^o] = \\ 
\nonumber 
 \left[ \frac{1}{4\pi \alpha a_o^2} \right] \int_0^{
\epsilon_0} \sigma_{PI} (2s^22p^5 7f \ ^3D_1) d\epsilon,
\end{eqnarray}
where $\epsilon$ is the energy relative to ionization threshold and up
to $\epsilon_o$ = $\sim$80 Ry.
An integration over the range shown in Fig.~4 yields the {\it partial} resonance
oscillator strength including all of the PECs and the non-PEC resonances
due to coupling of all 60 \fexviii levels. The sum over the oscillator
strengths corresponding to the 30 PECs gives a total $f_{PEC}$ = 2.31.
The integrated resonance oscillator strength $f_r$ is found to be 4.38.
In other words, the 30 PEC resonances involving transitions up to 
the \en = 3 levels of the core ion \fexviii  contribute over half of all the
continuum bound-free oscillator strength in photoionization of {\it any}
excited state of \fexvii. We note that, without loss of generality,
we chose an excited level to demonstrate the quantitative effect of PEC 
resonances; they manifest themselves in photoionization of most levels.

\subsection{Monochromatic Opacities}

Opacity calculations are a complex undertaking that requires atomic data
for a large number of ions in varying plasma environments \cite{symp,opal}.
One of the main reasons for the large-scale calculations carried out 
under the present study is to enable test calculations to benchmark 
available opacities for Fe ions. To that end, the present BPRM
calculations employ similar cut-offs, $n \leq 10$ and $\ell \leq n-1$, 
as the OP work \cite{op}. However, one usually expects only the ground 
state and low-lying metastable states to be significantly populated. The 
ionization fractions and level populations are computed using an 
equation-of-state, such as the modified Boltzmann-Saha formulation in the 
`chemical picture' \cite{mhd}, based on the premise that isolated atoms
exist, albeit perturbed by the plasma environment \cite{symp}. At low 
densities and temperatures the ion fractions and occupation probabilties 
of high-lying levels are several orders of magnitude smaller than those 
for the ground state and metastable levels. But in the high 
temperature-density regime N$_e > 10^{24}$ cm$^{-3}$, T$_e > 10^6$ K, 
approaching those in  stellar cores, electron-ion recombination rates can 
be large, and increase rapidly as N$_e^2$. Even a small population in 
excited levels would then be susceptible to the resonant enhancements 
due to PEC resonances, which are currently neglected and the cross 
sections for excited levels are taken to be nearly hydrogenic, instead of 
the accurate form exemplified in Figs. 2 and 3.

While we have discussed integrated resonances oscillator strengths 
embedded in bound-free cross sections, there is no {\em direct} 
equivalence or 1-1 correspondence with bound-bound oscillator strengths, 
as generally computed in opacities calculations. Among the factors that 
distinguish the two are overlapping profiles and large energy widths of 
PEC resonances, reflecting the coupling of continua belonging to many 
target levels and strong dipole moments among them. Owing to the huge 
scale of data needed, definitive checks can only be made by calculating 
revised opacities using atomic data, as outlined herein.   

A practical problem likely to be encountered is the high energy resolution 
needed to represent resonance profiles. Whereas scattering cross sections 
are bounded by the unitarity condition (viz.  \cite{pn11}, no such upper 
bound exists for individual values of photoionization cross sections. 
The peak photoionization values may rise arbitrarily high, and numerical 
integration would tend to be inaccurate as the resonance profile 
approaches a Delta function; the integral is finite but the width is 
extremely narrow and impractical to resolve. However, the statistical 
methodology adopted in opacities work is to employ the {\it opacity 
sampling} technique. Monochromatic opacity spectra are sampled at 
approximately 10,000 points (viz. \cite{op}), although the atomic data 
are much more finely resolved. It has been verified that the statistical 
averages of the most important quantity, the Rosseland Mean Opacity 
(RMO), do not deviate by more than 1-3\% even if the atomic cross 
sections are `sampled' at 10$^5$ or 10$^6$ points \cite{symp}. Therefore, 
the energy mesh of $\sim$30,000 points used in this work, predominantly 
in the region of covered by the high energy \en = 3 resonances, should 
suffice for accurate opacities calculations.

   \begin{figure}
   \centering
\resizebox{90mm}{!}
{ \includegraphics{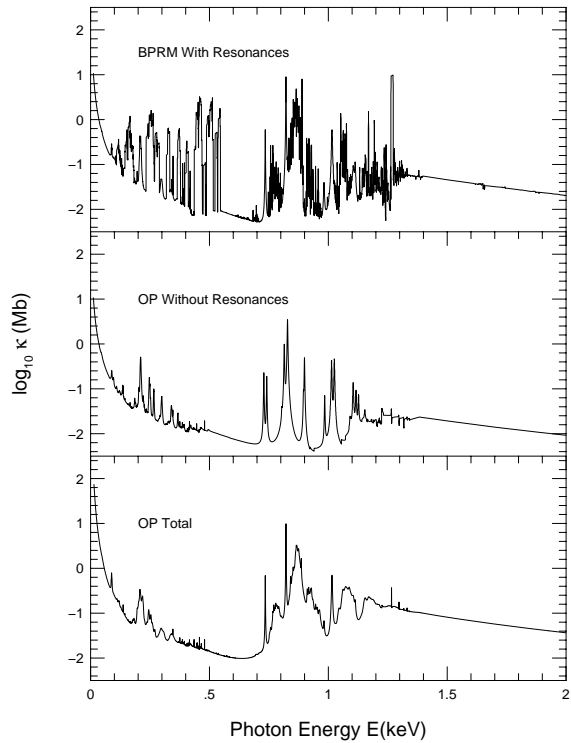}}
   \caption{Partial monochromatic opacity of \fexvii: $log_{10} \kappa$ (Mb)
at temperature T = 2.24 $\times 10^6$ K and electron density $N_e = 
10^{23}$ cm$^{-3}$, corresponding to the base of the solar convection 
zone where the \fexvii ion is the largest contributor to opacity. The 
top panel presents results using BPRM bound-free cross sections with 
resonances computed in this work, and the middle panel without resonances
using earlier data from the Opacity Project (OP). Identical datasets 
from OP are employed for the bound-bound transitions, therefore the 
differences are mainly due to the resonances included in the present work. 
For comparison, the bottom panel shows total OP opacities \cite{top} that 
include core-excitation resonances as lines, with autoionization widths
considered perturbatively, as well as the high-energy K-shell continuum 
opacity not yet included in BPRM calculations (see text).}
              \label{FigOP}%
   \end{figure}
Fig~5 presents the monochromatic opacity $\kappa$({\rm Fe~{\sc xvii}})
computed using all of the bound-free data in the present work, 454 
photoionization cross sections resolved as mentioned above. The
calculations are carried out using a newly developed code for
high-precision opacities, adapted from the earlier OP work \cite{symp}
(the OP opacities are available from the on-line server called OPserver
at the Ohio Supercomputer Center \cite{top}). The new code also employs 
a frequency mesh of $10^5$ points, an order of magnitude finer mesh than 
OP or OPAL, thus obviating some resolution issues in the monochromatic 
opacity spectra \cite{top}. Essential components of opacities calculations 
related to bound-bound transitions are retained as in the OP code: 
electron-impact and Stark broadening, free-free scattering, and 
electron-photon interaction in the Rayleigh, Thomson, and Compton 
scattering limits. However, resonance profiles have not yet been broadened. 
Whereas bound-bound line shapes are symmetrical, the giant PEC resonance 
profiles span hundreds of eVs (viz. Fig. 3 and 4)  and are asymmetric.
Theoretical formulations for broadening of resonances and algorithms are
still being developed. However, as mentioned, plasma effects in the
chemical picture \cite{mhd} are included a posteori, and do not affect 
the accuray of the atomic parameters computed herein. The details of the 
new opacities code and results will be reported later. But we note in 
passing that since the resonances would dissolve more readily than lines, 
it is likely to significantly affect {\it continuum lowering}, manfiest 
in opacity spectra, and highly senstiive to temperature and density.
Lines dissolve and eventually merge into the continuum in high 
temperature-density regime. Then a precise accounting may not be 
necessary; they are often treated as "unresolved transition arrays" 
(UTA) since J-J' transitions are merged together and often subsumed by 
line broadening \cite{roz}. However, 
our aim is to focus on delineation of atomic features as fully as possible 
so that their contributions to opacity in different temeperature-density 
regimes can be accurately ascertained.

The BPRM opacity cross sections in Fig.~5 are compared to two other sets 
of results from OP. Although over 20,000 oscillator strengths for 
bound-bound transitions are also computed, we use exactly the same data 
for lines as the earlier OP calculations described in \cite{symp}. Thus 
the differences between the the 60CC BPRM results from the present work 
in the upper panel of Fig.~5, and the limited OP without resonances in 
the middle panel, are entirely due to the differences in the bound-free 
data sets. The opacity calculations are done at a temperature-density 
representative of the plasma conditions at the base of the solar 
convection zone: log$_{10}$ T (K) = 6.35 and log$_{10}$ N$_e$ = 23. These parameters 
also lie in the range currently under investigation in the Z-pinch plasmas 
for measurment of transmission spectra \cite{bailey}. In addition to 
resonance contributions, there are some other differences. The OP data 
include extrapolated cross sections out to very high energies, $\sim$ 
500 Ry. The BPRM data have also been processed to include these 
high-energy "tails", but the form is slightly different. The background 
opacity is important to obtain a value for RMOs that spans over 4 decades 
in temperature, log$_{10}$ T(K) = 3.5 - 7.5. 

Calculations for the monochromatic opacity $\kappa_\nu$ are carried out
for each ion along isotherms in log$_{10}$ T for a range of electron
densities log$_{10}$ N$_e$. 
The Rosseland Mean Opacity (RMO) $\kappa_R$ is defined 
in terms of $\kappa_\nu$ as
\begin{equation}
 \frac {1}{\kappa_R} = \frac{\int_0^\infty g(u) \frac{1}{\kappa_\nu}
du}{\int_0^\infty g(u) du}, \ \ \ g(u) = u^4 e^{-u} (1 - e^{-u})^{-2},
\end{equation}
where $g(u)$ is the Planck weighting function (corrected for stimulated
emission). The $\kappa_\nu$ is primarily a function of the oscillator 
strengths $f$, photoionization cross sections $\sigma_\nu$, level 
populations $N_i$, and the line profile factor $\phi_\nu$,
\begin{equation}
\kappa^{bb}_\nu (i \rightarrow j) = \left( \frac{\pi e^2}{m_e c} \right)
N_i
f_{ij} \phi_\nu, \ \ \ \ \kappa^{bf}_\nu = N_i \sigma_\nu.
\end{equation}

Whereas further code developments are required to include all opacity 
contributions, we already find large enhancements in $\kappa_R$ due to 
resonances, primarily from the \en = 3 complex. The resonance 
contribution is included in existing opaciites codes as inner-shell 
bound-bound transitions\cite{note1}.  The BPRM value of $\kappa_R$, 
also including the bound-bound oscillator strengths computed in this 
work (as opposed to OP) yields a value of 223.8 cm$^2$/g. Using the same 
bound-bound data as OP, the BPRM $\kappa$-value is still 200.3 cm$^2$/g, 
compared to the OP value of 109.7 cm$^2$/g (Fig.~5). 
The bottom panel in Fig.~5 is the total monochromatic 
opacity spectrum of \fexvii from OP, including all contributions, with 
the total RMO $\kappa_R$ = 306.9 cm$^2$/g. Thus the BPRM value using the 
data computed in this work is 27\% lower. This is primarily because of 
two factors:
(I) The high-energy "tails" are to made more precise with even more extended
BPRM calculations including the K-ionization thresholds, which would
attenuate the bound-free opacity in the large energy range between the L
and the K levels. Although the K-shell opacity is not too significant,
the K-shell resonances would thereby be included. (II) The bound-bound
oscillator strengths are computed up to \en = 10. Therefore, there is a
relatively small gap in cross sections in
the region between \en~= ~10~-~$\infty$, below the
photoionization thresholds of the 454 bound levels. In the earlier OP
work, we began the tabulation of cross sections at E = $-z^2/\nu^2$
($\nu \sim$ 10) below each threshold. Employing a similar approximation
increases the BPRM RMO value from 223.8 cm$^2$/g to 260.7 cm$^2$/g, to
within 20\% of the total OP RMO. Owing
to the signficance of (I) and (II), we plan to carry out more extensive
calculations than in earlier works, as well investigate if higher-\en
resonance complexes with \en = 4 or 5 might be needed.

\subsection{Unified Electron-Ion Recombination}

One of the important charateristics of PEC resonances is that they entail
excited states with valence electrons that are weakly bound to the core
ion. Therefore, during core excitation the outer electron remains 
essentially a `spectator', temporarily attached to the excited core level, 
and autoionizes as that level decays to the ground state. The analogy and
connection between the PEC resonances and the dielectronic recombination 
process is well known \cite{ys87,np94,pn11}. 
It is expected that the total electron-ion 
recombination rates of \fexvii will also be commensurately enhanced by 
inclusion of the PEC resonances via detailed balance (Milne relation, 
e.g. \cite{np94,pn11}).

A major application of the computed cross sections is in benchmarking 
total unified recombination rate coefficients, including radiative 
and dielectronic recombination, for \fexvii at {\em high temperatures}, 
including those where \fexvii is abundant in coronal plasmas, such as in the 
solar corona and solar flares, up to at least $10^7$K. That is possible because 
we have considered a large energy range up to the excitation of 
$n$ = 3 levels of the 
recombining ion \fexviii. Towards that end, we have also repeated the entire 
photoionization calculation to obtain partial cross sections for the 
454 bound levels of \fexvii with the residual ion in the ground state 
alone.

\section{Conclusion}

In this report we have presented results from a pilot project of complete
BPRM calculations for photoionization of an atomic system, Fe~XVII, larger 
than a He- or Li-like ion, (h$\nu$ + Fe~XVII $\rightarrow$ e + Fe~XVIII). 
The aim was to study in detail the extent and range of high energy 
resonances of importance in practical applications. 
The BPRM calculations consider all fine structure levels up to $n(SLJ) 
\leq 10$, with spectroscopic identification. In addition to 
photoionization cross sections, the bound-bound oscillator strengths for 
transitions among the 454 computed levels of \fexvii are also being 
computed. These datasets of atomic parameters should be of unprecedented 
accuracy and generally useful.

The comprehensive calculations were carried out so as to compare with 
the erstwhile OP calculations that treated resonances as lines. A clear 
distinction is made between pure bound-bound transitions and bound-free 
transitions into autoionizing levels. The results from this work 
demonstrate that opacities may be computed using BPRM cross sections and 
transition probabilities, and it is likely that plasma opacities in 
general, and those of Fe ions in particular, would be different from 
earlier ones using atomic cross sections that accurately consider the
energy distribution of resonances. 

Mnochromatic opacities of \fexvii are computed and compared with the OP
work. These are not the final opaciites, and some further developments
are still necessary, such as resonance broadening mechanisms and K-shell
contributions. Nevertheless, we are able to compute complete BPRM 
radiative datasets and Rosseland mean opacities that are sufficiently 
close to OP values to imply that future calculations of plasma opaciites 
can be carried out with higher precision. 

Three other applications of the present work might be pointed out. (i)
Calculation of accurate unified electron-ion recombination rate 
coefficients. (ii) Benchmarking experimental measurements of absolute
photoionization cross sections on accelerator based light sources, that
are now being made for multiply charged Fe ions (e.g. \cite{hassan}). 
It has now been established that the experimental beams contain ions not 
only in the ground state but also in several metastable levels. An 
admixture of ground state plus a few excited metastable levels is 
therefore necessary to benchmark experimental measurements against theory. 
In addition, the experimental measurements are capable of reaching 
energies where PEC resonances occur (e.g.  \cite{hetal}). (iii) The 
radiative data for \fexvii should be useful for X-ray spectral 
diagnostics and non-LTE models ((Ref.  \cite{cpe03} presents a detailed 
Grotrian diagram of \fexvii levels up to $n$=4). 

Estimates of uncertainties in the large sets of parameters reported in 
this paper, as well as that required in future calculations of opacities, 
are as follows. The accuracy of the 454 theoretically computed energy 
levels has been ascertained in the text by comparison with experimentally 
observed levels, in general better than 1\%. The differences between BPRM 
cross sections and measurements for photoionization and recombination for 
\fexvii are found to be within experimental uncertainties of $\sim$10-20\% 
\cite{znp}. Similarly, most of the strong transition in the extensive 
dataset of \fexvii oscillator strengths should be better than 10\% accuracy 
\cite{n03}. The present calculations for photoionization employ a much 
larger 60CC eigenfunction expansion than the 3CC calculations in previous 
studies \cite{znp,n03}, and are expected to be more accurate. 
Uncertainites in the BPRM monochromatic opacites should be commensurate 
with those in the underlying 60CC data. However, the final accuracy of 
derived mean Rosseland and Planck opacities ions is as yet undetermined 
since photoionization cross sections extrapolated or computed at high 
energies are to be included for all 454 bound states. The problem is 
further complicated since the aim is to obtain statistical averages with 
sampled opacities (albeit on a finer mesh of 10$^5$ frequencies as 
opposed to 10$^4$ in earlier calculations) that would approach the 
accuracy, $\sim$1\%, needed to improve over existing uncertainties of 
$\sim$5\% \cite{sb04}. 

Electronic files for photoionization cross sections, energy levels, and
oscillator strengths are available electronically from the NORAD website:
www.astronomy.ohio-state.edu/$\sim$nahar/nahar$\_$radiativeatomicdata

\begin{acknowledgments}
This work was partially supported by the NASA Astronomy and Physics
Research Analysis Program (SNN) and the U.S. Department of Energy (AKP).
The computational work was carried out at the Ohio Supercomputer Center
in Columbus Ohio.
\end{acknowledgments}

%
 
%
\def\amp{{Adv. At. Molec. Phys.}\ }
\def\apj{{ Astrophys. J.}\ }
\def\aap{Astronomy and Astrophysics}
\def\apjs{{Astrophys. J. Suppl.}\ }
\def\apjl{{Astrophys. J. (Lett.)}\ }
\def\aj{{Astron. J.}\ }
\def\aa{{Astron. Astrophys.}\ }
\def\aas{{Astron. Astrophys. Suppl.}\ }
\def\adndt{{At. Data Nucl. Data Tables}\ }
\def\cpc{{Comput. Phys. Commun.}\ }
\def\jqsrt{{J. Quant. Spectrosc. Radiat. Transf.}\ }
\def\jpb{{J. Phys. B}\ }
\def\pasp{{Pub. Astron. Soc. Pacific}\ }
\def\mn{{Mon. Not. R. Astron. Soc.}\ }
\def\mnras{{Mon. Not. R. Astron. Soc.}\ }
\def\MNRAS{{Mon. Not. R. Astron. Soc.} }
\def\pra{{Phys. Rev. A}\ }
\def\ps{{Phys. Scr.}\ }
\def\prl{{Phys. Rev. Lett.}\ }
\def\zpds{{Z. Phys. D Suppl.}\ }


\begin{thebibliography}{10}
\bibitem{note1} Opacities sources do not usually report monochromatic
opacity spectra for individual ions, viz. \cite{top}. Code development
is under way to isolate different opacity contributions in the OP work
to provide detailed comparisons with improved opacities, using the BPRM
data exemplified by the pilot calculations presented herein.
\end{thebibliography}
\end{document}